\documentclass[10pt,twocolumn,letterpaper]{article}

\usepackage[pagenumbers]{cvpr} %
\usepackage{graphicx}
\RequirePackage[shortlabels,inline]{enumitem}
\setlist[itemize]{noitemsep,leftmargin=*,topsep=0em}
\setlist[enumerate]{noitemsep,leftmargin=*,topsep=0em}

\definecolor{cvprblue}{rgb}{0.21,0.49,0.74}
\usepackage[pagebackref,breaklinks,colorlinks,allcolors=cvprblue]{hyperref}
\usepackage{textcomp}
\usepackage{tabularray}
\usepackage{graphicx}
\usepackage{float}
\restylefloat{table}
\usepackage{mathtools}

\title{$\rho$-NeRF: Leveraging Attenuation Priors in Neural Radiance Field for 3D Computed Tomography Reconstruction}

\author{Li Zhou,
Changsheng Fang,
Bahareh Morovati,
Yongtong Liu,
Shuo Han,
Yongshun Xu\thanks{Dr. Xu is now with Samsung NeuroLogica and his work was done in the University of Massachusetts Lowell.},
Hengyong Yu\thanks{Corresponding author, email: hengyong-yu@ieee.org. }\\
\\
University of Massachusetts Lowell\\
}

\begin{document}
\maketitle
\begin{abstract}
This paper introduces~$\rho$-NeRF, a self-supervised approach that sets a new standard in novel view synthesis (NVS) and computed tomography (CT) reconstruction by modeling a continuous volumetric radiance field enriched with physics-based attenuation priors. The ~$\rho$-NeRF represents a three-dimensional (3D) volume through a fully-connected neural network that takes a single continuous four-dimensional (4D) coordinate—spatial location~$(x, y, z)$ and an initialized attenuation value~$(\rho)$—and outputs the attenuation coefficient at that position. By querying these 4D coordinates along X-ray paths, the classic forward projection technique is applied to integrate attenuation data across the 3D space. By matching and refining pre-initialized attenuation values derived from traditional reconstruction algorithms like Feldkamp-Davis-Kress algorithm (FDK) or conjugate gradient least squares (CGLS), the enriched schema delivers superior fidelity in both projection synthesis and image reconstruction, with negligible extra computational overhead. The paper details the optimization of~$\rho$-NeRF for accurate NVS and high-quality CT reconstruction from a limited number of projections, setting a new standard for sparse-view CT applications.
\end{abstract}
    
\section{Introduction}
\label{sec:intro}

\begin{figure*}
  \centering
   \includegraphics[width=0.9\linewidth]{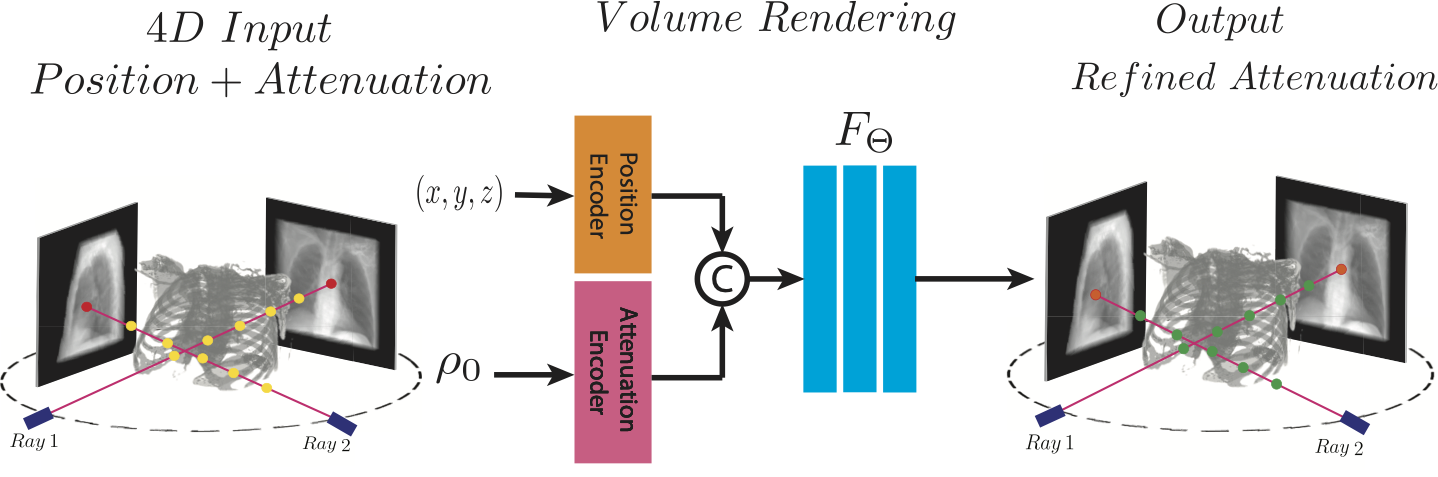}
   \caption{Overview of the~$\rho$-NeRF framework for CT reconstruction. Discrete X-ray sample points integrate attenuation priors $\rho_0$ with spatial coordinates $(x, y, z)$. Attenuation values are updated along the X-ray paths by applying forward projection technique with encoders and the mapping function~$F_{\Theta}$.}
   \label{fig:over}
\end{figure*}

X-ray imaging, a form of penetrative imaging to capture detailed internal structures, is widely applied across fields, such as medicine, industrial inspection, materials science, \emph{etc.}~\cite{maire2001application, singh2005advanced}. 
While high-resolution projections can produce high-quality CT images, they require extended exposure time, and excessive radiation pose health risks, particularly in medical applications. Consequently, one of the recent research topics focused on sparse-view and low-dose techniques to reduce radiation exposure and/or increase temporal resolution while maintaining diagnostic utility~\cite{zang2018super,morovati2023reduced,wang2024lomae}, particularly for the situations where there are no slip-ring in the CT scanner. However, in cases like 3D cone beam CT (CBCT) reconstruction, sparse-view data often limits resolution and introduces artifacts compared with full-view data, which highlights the need for advanced computational methods to accurately synthesize views and generate anatomical structure from a limited number of projections.

CT reconstruction techniques can be categorized into analytical, iterative, and hybrid data-driven approaches. Analytical methods, such as filtered backprojection (FBP)~\cite{herman2009fundamentals} and its cone-beam variants~\cite{feldkamp1984practical}, are effective with dense projection data but fall short under sparse-view conditions. Iterative approaches, including methods like simultaneous algebraic reconstruction technique (SART) family~\cite{gilbert1972iterative,andersen1984simultaneous} and total variation (TV)-minimization techniques~\cite{zeng2010medical}, address some of these limitations by optimizing the reconstruction through iterative updates. However, these methods require considerable computational resources and are not always practical for real-time or large-scale applications. Recently, hybrid data-driven approaches that incorporate deep learning have enhanced the traditional techniques by using neural networks to predict, fill gaps, and improve denoising for sparse data~\cite{morovati2024impact,han2024physics}. While these models achieve impressive results, they often require extensive labeled datasets for training, which limits their adaptability and generalizability to different datasets without further fine-tuning. Given these limitations, self-supervised methods offer a promising alternative by reducing reliance on labeled data and improving adaptability across various reconstruction tasks. Inspired by advancements in 3D reconstruction in computer vision, recent works~\cite{adler2018learned,zang2021intratomo,zha2022naf,cai2024structure} have explored the applications of self-supervised methods for CT reconstruction, where internal structures are captured through X-ray transmission imaging rather than reflective imaging.

In the computer vision field, 3D reconstruction typically represents shapes as discrete point clouds or meshes. Implicit neural representations (INRs) have become popular as they map discrete points to continuous functions, enabling smoother and more accurate modeling of complex geometries. NeRF~\cite{mildenhall2021nerf}, a leading model in this space, leverages INRs for reflective imaging with camera rays, mapping spatial position and viewing direction in 5D coordinates to RGB color and volume density, allowing photorealistic novel view synthesis (NVS) through volumetric rendering. However, NeRF and similar models~\cite{pumarola2021d,wang2021nerf,tancik2022block} are primarily designed for reflective imaging and can not be directly applied to X-ray imaging to capture internal structures. Recent INR-based approaches for CT reconstruction, such as neural attenuation field (NAF)~\cite{zha2022naf} and SAX-NeRF~\cite{cai2024structure}, adapt NeRF for X-rays by mapping 3D spatial coordinates to attenuation coefficients. While these adaptations improve reconstruction quality, they only consider spatial location, and further enhancements to accuracy come with advanced mapping functions/models which significantly increased computational costs.

To address the limitations of current methods, we present~$\rho$-NeRF framework (see ~\cref{fig:over}), a self-supervised approach designed for X-ray-based 3D tomographic reconstruction. We adopt the CBCT geometry as defined in TIGRE~\cite{biguri2016tigre}, making~$\rho$-NeRF adaptable across a range of CT imaging settings. Our key contributions are as follows:
\begin{itemize}[noitemsep,topsep=0pt]
    \item Redefining Input with Attenuation Priors: By examining the inherent properties of CT imaging, we redefine the input for implicit neural representations (INRs) as a 4D coordinate system consisting of spatial location $(x,y,z)$ and an initialized attenuation value~$(\rho)$. This enables~$\rho$-NeRF to model a continuous volumetric radiance field, capturing the 3D volume with a fully-connected neural network. By querying these 4D coordinates along X-ray paths and applying forward projection technique,~$\rho$-NeRF is compatible with other NeRF-based CT reconstruction frameworks and achieves enhanced fidelity in sparse-view applications.
    \item Efficient Attenuation Initialization: We introduce interpolation methods—mean, nearest neighbor, and trilinear—to initialize attenuation values~$(\rho)$ at a sparse spatial location from its 8 neighboring points on the grids. We use these techniques to match and initialize attenuation values, which are pre-initialized by the traditional algorithms such as FDK and CGLS. Also, a lightweight, learnable linear transformation serves as feature encoding for further refinement. Together, these techniques enhance projection synthesis and image reconstruction with minimal computational cost.
    \item Benchmark Validation on X3D Dataset: Extensive evaluations on the X3D dataset, a large-scale benchmark for X-ray 3D reconstruction, confirm~$\rho$-NeRF’s state-of-the-art performance in both NVS and CT reconstruction across different X-ray applications. This 4D scene representation captures high-resolution geometry and enables high-fidelity anatomical reconstructions, establishing a new benchmark for higher accuracy and efficiency in sparse-view CT applications.
\end{itemize}
In summary,~$\rho$-NeRF leverages attenuation priors to address the limitations of existing INR-based CT reconstruction methods, achieving notable improvements in performance and fidelity with negligible extra computational demands. This work sets a new benchmark for low-dose, sparse-view CT imaging, advancing the field toward efficient, high-quality 3D X-ray reconstructions.

\section{Related Work}
\label{sec:relatedwork}

\subsection{3D CT Reconstruction}
CT reconstruction methods are broadly categorized into analytical, iterative, and hybrid data-driven approaches. 
Foundational analytical algorithms, such as filtered backprojection (FBP)~\cite{herman2009fundamentals} and its cone-beam extension, Feldkamp-Davis-Kress (FDK)~\cite{feldkamp1984practical}, reconstruct attenuation coefficients from projections by solving the Radon transform. Variants with different filters and Parker weights improve image quality under dense data but produce artifacts under sparse-view conditions. 
Iterative algorithms address these limitations by framing reconstruction as a maximum a posteriori (MAP) problem, solved through iterative optimization. 
Gradient-based approaches like the algebraic reconstruction technique (ART) family (SART~\cite{andersen1984simultaneous}, SIRT~\cite{gilbert1972iterative}, and OS-SART~\cite{zeng2010medical}) vary in update strategy:   SART and SIRT update the full projection set at once, yielding high quality images with high computational cost, and OS-SART strikes a balance with efficient and subset-based updates. 
Total variation (TV) minimization based methods (\emph{e.g.}, ASD-POCS~\cite{defrise2006truncated}, OS-ASD-POCS~\cite{zeng2010medical}, and AwASD-POCS~\cite{sidky2008image}) iteratively refine reconstructions to suppress noise, with variants improving computational efficiency or edge preservation. 
Krylov subspace algorithms (\emph{e.g.}, CGLS~\cite{dabravolski2014dynamic}, LSQR~\cite{flores2016application}) achieve faster convergence by focusing on the eigenvectors of the residual in descending order, which allows for increased convergence rates compared to the SART family. While the SART-based methods are effective with high-quality projection data, Krylov subspace algorithms, like CGLS, are well-suited for handling large datasets or low-quality data, offering both speed and robustness, especially in CT denoising.

Hybrid data-driven approaches combine the traditional methods with deep learning. These methods enhance reconstruction by using neural networks to 1) predict and fill gaps in limit-view or sparse-view projections~\cite{anirudh2018lose, li2019promising, jiang2021reconstruction}, 2) directly infer attenuation coefficients as denoising tasks from training data~\cite{lahiri2023sparse, zhou2024gradient, wang2024lomae}, and 3) optimize differentiable processes to accelerate computations~\cite{shen2018intelligent, yu2010fast, xu2020sparse}. While these models achieve impressive results and require minimal data during inference, they depend on extensive, domain-specific datasets for training, which restricts their adaptability to different applications without further fine-tuning. Consequently, hybrid data-driven models excel in specific contexts but face challenges in generalizability across different CT reconstruction scenarios. 
Given these limitations, studying self-supervised methods could provide a promising alternative by reducing reliance on labeled data and improving adaptability across different reconstruction tasks.

\subsection{Implicit Neural Representation and Rendering}
Learning implicit neural representations (INRs) has become a popular approach in 3D scene reconstruction, offering a way to transform discrete data points into continuous functions for learning-based 3D geometry modeling~\cite{genova2019learning,mescheder2019occupancy,park2019deepsdf}.
Neural rendering leverages INRs to map discrete data to coordinate-based continuous representations, typically using implicit functions parameterized by neural networks~\cite{sitzmann2019deepvoxels, lombardi2019neural, mildenhall2021nerf}. One prominent method is the neural radiance field (NeRF)~\cite{mildenhall2021nerf}, a model that has set a standard for high-quality novel view synthesis.

For reflective imaging, NeRF models a scene by mapping spatial position~$(x, y, z)$ and viewing direction~$(\theta,\phi)$ along camera rays to RGB color~$(r,g,b)$ and volume density~$(\sigma)$. This allows NeRF to produce high-quality novel views by integrating color and densities along rays through volumetric rendering~\cite{kajiya1984ray}.
Numerous works have aimed to improve NeRF’s efficiency~\cite{barron2021mip, barron2022mip,chen2022tensorf,chen2023neurbf}, and extend its application scope, such as generative modeling, unbounded scenes, and RGB-D synthesis. More recently, INR-based methods have been explored for CT image reconstruction~\cite{zha2022naf,zhang2023dynamic,cai2024structure}, transforming discrete samples into continuous representations of internal structures.

For X-ray imaging, adapting NeRF is necessary to address fundamental differences from reflective imaging. Unlike visible light, which reveals surface color and reflectance, X-ray penetrates objects to capture internal structures through attenuation. Neural attenuation fields (NAF)~\cite{zha2022naf} modify the NeRF framework for 3D CBCT reconstruction by mapping spatial positions~$(x,y,z)$ directly to attenuation coefficients~$(\rho)$. 
Follow-up works~\cite{chen2023cunerf,hu2024umednerf,shao2024dynamic,liu2024geometry} refined this method by introducing advanced modeling functions or complex encoding techniques, such as Transformer-based network in SAX-NeRF~\cite{cai2024structure}. However, these adaptations often increase computational demands.
To address this issue, we introduce attenuation priors to INR-based CT reconstruction algorithms, achieving performance improvement with minimal computational overhead.

\section{Method}
\label{sec:method}

\subsection{X-ray Sampling}
\label{sec:ray}
With appropriate preprocessing steps, the measurement of X-ray attenuation can be approximately modeled by a linear integral, represented as:
    \begin{equation}
      I(\mathbf{r})=\int_{t^{n}}^{t^{f}}\rho (\mathbf{r}(t)) \, dt,
      \label{eq:beer}
    \end{equation}
    where~$\rho(\mathbf{r}(t))$ represents the attenuation coefficient at each point $\mathbf{r}(t)$ parametrized by $t$ along the path, and ~$t_{n}$ and~$t_{f}$ are the near and far limits to account for the material’s effective attenuation coefficients along the ray path. 
To align with the voxel grid of view, the ray is divided into evenly spaced bins within the near and far limits, with one point uniformly sampled within each bin. \cref{eq:beer} can be discretized as:
    \begin{equation}
      I(\mathbf{r})=\sum_{j=1}^{M}\rho_{j} \, \Delta t,
      \label{eq:beer_discrete}
    \end{equation}
where~$M$ is number of points sampled along the ray,~$\rho_{j}$ denotes the attenuation coefficient at the each sampled bin paramterized by $t_j$, and $\Delta t$ is the sampling interval length. 
The objective of tomographic reconstruction is to estimate the distribution of $\rho$ and produce it as a discrete volume, using X-ray projections captured from $N$ different angles.

\begin{figure}[t]
  \centering
   \includegraphics[width=0.9\linewidth]{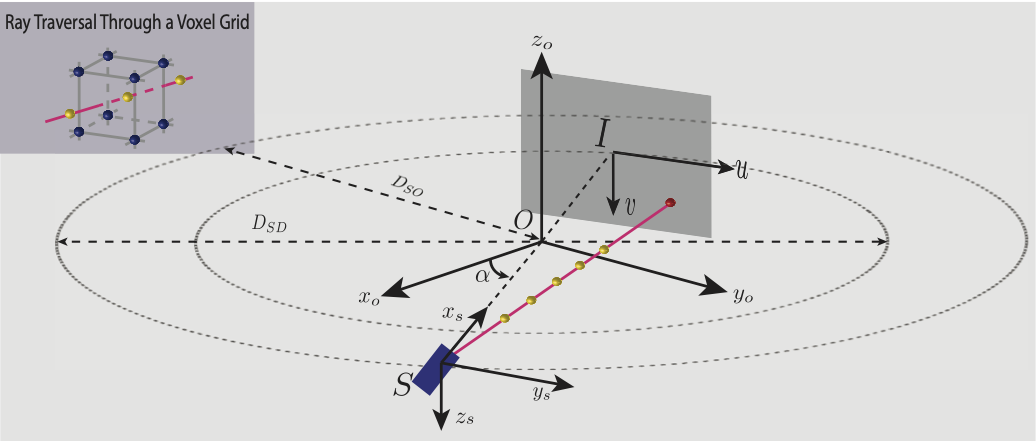}
   \caption{Geometric configuration of CBCT and X-ray sampling.}
   \label{fig:geometry}
\end{figure}

As shown in~\cref{fig:geometry}, we model the CBCT geometry by following the conventional definitions and incorporating a local coordinate system. An X-ray source~$S$ rotates around the object along a circular trajectory defined by the rotation angle~$\alpha$. Unlike reflective imaging setups, the object is centered at the origin~$O$ of a world coordinate system, and the projections are captured on a flat panel detector positioned opposite to the source. The imaging object is represented as a cube and discretized into voxel units for detailed analysis. This detector records X-ray projections in its own local coordinate system, \emph{i.e.} an image coordinate system.

Let~$\mathbb{I}=\left\{ I_{i}\right\}_{i=1,...,N}$ be the set of X-ray projections obtained from~$N$ different rotation angles. Each projection~$I_i$ records the attenuation of X-rays as they traverse the object. We model an X-ray path as~$\mathbf{r}(t)=\mathbf{o}_{o}+t\mathbf{d}_{o}\in \mathbb{R}^{3}$, bounded by the near and far limits~$t_{n}$ and~$t_{f}$. The direction of each ray is encoded by a 3D unit Cartesian  vector~$\mathbf{d}_{o}$ and X-ray source position $\mathbf{o}_{o}$ derived from the geometric configuration between the X-ray source and the detector. Specifically,~$\mathbf{d}_{o}$ and $\mathbf{o}_{o}$ are calculated based on the rotation angle~$\alpha$, the distance from source to detector ($D_{SD}$), and the distance from source to object ($D_{SO}$). 
At the position of source point $\mathbf{o}_{o}$ 
\begin{equation}
    \textbf{o}_{o}=\left ( D_{SO} \cdot cos(\alpha ), D_{SO}\cdot sin(\alpha ), 0\right ),
  \label{eq:ray2}
\end{equation}
we define a local coordinate system $(x_s,y_s,z_s)$ along three unit vectors $(-cos(\alpha),-sin(\alpha ),0),(-sin(\alpha ),cos(\alpha ),0)$ and $(0,0,-1)$. We assume that the detector is perpendicular to $x_s$-axis. Considering a local detector coordinator system ~$(u,v)$ with $u$-axis parallel to the $x_oy_o$-plane and $y_s$-axis and $v$-axis parallel to $z_o$-axis and $z_s$, $(0,0)$ is the projection position of the X-ray source.  
The transformation of the ray direction within the global coordinate system can be expressed through an affine transformation:
\begin{equation}
    \mathbf{d}_{I} =\frac{1}{\sqrt{u^2+v^2+(D_{SD})^2}}\left(-D_{SD}, u, -v\right), \\
    \mathbf{d}_{o}=R \cdot \mathbf{d}_{I},
  \label{eq:ray1}
\end{equation}
where~$\mathbf{d}_{I}$ is ray's direction for ~$\alpha=0$, and~$R$ is the rotation matrix associated with the source rotation angle ~$\alpha$  

 \begin{equation}
 R=\left(\begin{smallmatrix}
  cos(\alpha)  &  sin(\alpha) &0 \\
  -sin(\alpha) &  cos(\alpha )  & 0 \\
  0& 0 & 1 \\
 \end{smallmatrix}\right).
   \label{eq:rotation}
 \end{equation}

\subsection{Attenuation Priors Scene Representation}
We follow the conventional idea of NeRF~\cite{mildenhall2021nerf}, adapting it for X-ray imaging by using attenuation values instead of color and density. The following pipeline, illustrated in~\cref{fig:over}, details our approach to integrate attenuation priors into a neural radiance field.
    
    \textbf{Attenuation Modeling:} 
We model tomographic images as a continuous 4D attenuation field, represented as a function:
\begin{equation}
    F_{\Theta} : (x, y, z, {\rho}_{0}) \rightarrow (\rho),
  \label{eq:map}
\end{equation}
where~$(x,y,z)$ is a 3D spatial coordinate,~$\rho_{0}$ represents an initialized attenuation value, and~$\rho$ denotes the learned attenuation coefficient. The neural network, parameterized by~$\Theta$, is optimized to map each 4D input to its corresponding attenuation value, yielding a continuous representation of the attenuation field.

\textbf{Attenuation Initialization:} 
The initialization of attenuation priors begins with the traditional reconstruction algorithms, such as FDK and CGLS, to produce initial attenuation estimates,~$\rho_{0}$, for the 3D attenuation map in the global coordinate system. While these voxel-based initial values provide a structured foundation, further refinement is needed to achieve a continuous, high-fidelity reconstruction. To this end, we employ three interpolation methods—mean, nearest neighbor, and trilinear—to smooth and adapt attenuation values at points where X-ray paths intersect the voxel grid, as shown in~\cref{fig:geometry}. The attenuation value at a sampled point is determined using 8 neighboring voxel vertices via their mean, nearest neighbor, or trilinear interpolation. This initialization pipeline offers robust priors that support accurate projection synthesis and high-quality reconstruction from any view.

\textbf{Feature Encoding:} 
Previous studies~\cite{hornik1989multilayer, rahaman2019spectral} have shown that deep networks tend to favor low-frequency representations, limiting their ability to capture fine details in color and geometry. By mapping low-dimensional inputs into a higher-dimensional space with high-frequency encodings, neural networks better capture these high-frequency variations. Given that X-ray imaging naturally features homogeneous tissue regions with sharp boundaries at anatomical transitions, resulting low variation of the image, we adopt a learning-based encoding mechanism to leverage these properties. For spatial positions, we use a~\textit{hash encoder}~\cite{muller2022instant}—a sparse, learning-based encoding method—to efficiently represent position details. For attenuation values, we apply a lightweight~\textit{linear transformation} to refine the input for enhanced accuracy in reconstruction.

\textbf{Model Optimization:} 
The optimization process of ~$\rho$-NeRF involves minimizing the difference between the predicted and actual X-ray projections across the dataset. The network is trained to predict attenuation coefficients for each 4D input (spatial position and initial attenuation) by using an objective function based on the mean squared error (MSE) between the synthesized projections and the ground truth:
\begin{equation}
Loss(\Phi,\Psi,\mathrm{A}) = \sum_{\textbf{r}\in \textbf{R}}\left\| I(\textbf{r})-\hat{I}(\textbf{r})\right\|^{2},
  \label{eq:loss}
\end{equation}
where~$I(\textbf{r})$ and~$\hat{I}(\textbf{r})$ are real and synthesized projections from ray~$\textbf{r}$ respectively, and~$\bold{R}$ is a batch of rays.~$\Phi$ is the learned hash encoder for spatial location~$(x,y,z)$,~$\Psi$ is the learned linear encoder for initial attenuation~$\rho_{0}$, and~$\mathrm{A}$ is the attenuation coefficients network. $\hat{I}(\textbf{r})$ is computed by Eq.~\eqref{eq:beer_discrete} where $\rho_{j}$ is estimated by the network~$\mathrm{A}$ and the related encoders ~$\Phi$ and ~$\Psi$. This is implemented by the classic forward projection technique that corresponds to the rendering model in the computer vision field. The CT reconstruction process is achieved by feeding the voxel grid coordinates, sized to the desired dimensions, into the trained neural function~$F_{\Theta}$ to predict the corresponding attenuation coefficients. 
In our work, we use the MLP model architecture from ~\cite{mildenhall2021nerf} as the foundation for our study. Additionally, we explore a transformer-based model that incorporates X-ray structural awareness~\cite{cai2024structure} to further validate the applicability of our proposed method. The final output is represented as a discrete 3D matrix.

\section{Experiments and Results}
\label{sec:experiments}

\begin{table*}[h!]
\centering
\begin{tblr}{
  vline{2} = {-}{},
  hline{1,7} = {-}{0.08em},
  hline{2} = {-}{0.05em},
}
         & FDK          & ASD-POCS     & CGLS         & NAF          & $\rho$-NAF & SAX-NeRF     & $\rho$-SAX-NeRF \\
Jaw      & 28.57/0.7816 & 33.23/0.9322 & 28.80/0.8536 & 34.08/0.9345 & 35.23/0.9512            & 35.30/0.9503 & \textbf{35.25/0.9522}                 \\
Foot     & 24.53/0.5999 & 29.98/0.9208 & 25.39/0.7901 & 31.76/0.9341 & \textbf{32.84/0.9506}            & 32.25/0.9403 & 32.72/0.9481                 \\
Chest    & 22.89/0.7861 & 31.13/0.9422 & 24.44/0.7721 & 32.65/0.9609 & \textbf{35.07/0.9733}            & 34.46/0.9725 & 34.81/0.9747                \\
Aneurism & 28.07/0.7295 & 34.71/0.9858 & 30.08/0.9272 & 38.02/0.9883 & 39.43/0.9928            & 41.46/0.9956 & \textbf{41.41/0.9966}                 \\
Bonsai   & 24.53/0.7275 & 32.71/0.9529 & 26.27/0.8200 & 34.07/0.9606 & 34.79/0.9705            & 36.11/0.9751 & \textbf{36.85/0.9798}                
\end{tblr}
\caption{Quantitative comparison of CT reconstruction from different methods in terms of 
 PSNR/SSIM, where ~\textbf{bold} fonts indicate the best results.}
\label{tab:reconstruction}
\end{table*}

\begin{table*}[h!]
\centering
\begin{tblr}{
  vline{2} = {-}{},
  hline{1,7} = {-}{0.08em},
  hline{2} = {-}{0.05em},
}
         & NAF          & $\rho$-NAF & SAX-NeRF     & $\rho$-SAX-NeRF \\
Jaw      & 40.10/0.9988  & 43.18/0.9991            & 42.89/0.9991 & \textbf{43.37/0.9991}                 \\
Foot     & 38.13/0.9924 & 46.47/0.9994            & 46.64/0.9994 & \textbf{47.56/0.9994}                 \\
Chest    & 42.26/0.9993 & 47.62/0.9996            & 47.41/0.9994 & \textbf{47.66/0.9996}                \\
Aneurism & 41.03/0.9991 & 46.84/0.9996            & 52.90/0.9998 & \textbf{53.78/0.9998}                 \\
Bonsai   & 49.12/0.9990 & 51.64/0.9992            & 52.29/0.9995 & \textbf{53.44/0.9995}                
\end{tblr}
\caption{Quantitative comparison of novel view synthesis from different methods in terms of PSNR/SSIM, where ~\textbf{bold} fonts indicate the best results.}
\label{tab:projection}
\end{table*}

\subsection{Experimental Setup}

Following novel Nerf-based works~\cite{zha2022naf, cai2024structure}, we evaluate our method using publicly available human organ CT datasets, including LIDC-IDRI~\cite{armato2011lung} and scientific visualization dataset~\cite{data2022}.
Using the open-source tomographic toolbox TIGRE~\cite{biguri2016tigre}, we simulate 100 projections per case with 3\% noise over a 0°–180° range. For the novel view synthesis (NVS) task, we split these projections evenly, with 50 for training and 50 for testing. For sparse-view CT reconstruction, the CT volumes are reserved as ground truths for evaluation.

The $\rho$-NeRF framework is implemented in PyTorch and trained on an NVIDIA RTX 3090 GPU. We optimize the model using the Adam optimizer~$(\beta_{1}=0.9,\beta_{2}=0.999)$ for at most 3,000 iterations, with an inital learning rate of 0.001, reduced to 0.0001 halfway through training. For efficient processing, we use a batch size of 1,024 rays, with each ray sampled at 192 points to adequately represent the attenuation field across the CT volume. After training, we quantitatively evaluate the reconstruction by computing both peak signal-to-noise ratio (PSNR) and structural similarity index (SSIM) metrics.
PSNR (in dB) provides a statistical measure of artifact suppression performance, while SSIM assesses perceptual differences between two images. Higher PSNR and SSIM values indicate more accurate reconstructions.

\subsection{Main Results}
We include comparisons with the conventional CT reconstruction algorithms, including FDK (physics-based), AS-POCS (iterative), and CGLS (Krylov subspace), as well as the state-of-the-art NeRF-based methods, named NAF and SAX-NeRF. 
We deploy the~$\rho$-NeRF framework to NeRF-based models by utilizing their neural networks as the mapping function~$F_{\Theta}$, naming the enhanced models $\rho$-NAF and $\rho$-SAX-NeRF. These new models integrate attenuation priors, the nearest neighbor interpolation method, and the two proposed feature encoders, demonstrating the effectiveness and adaptability of our approach. Unless otherwise noted, this configuration is used throughout the experiments. A detailed analysis of these design choices is provided in~\cref{sec:ablation}, with additional quantitative and qualitative results available in the supplementary materials.

\begin{figure*}[h!]
  \centering
   \includegraphics[width=0.9\linewidth]{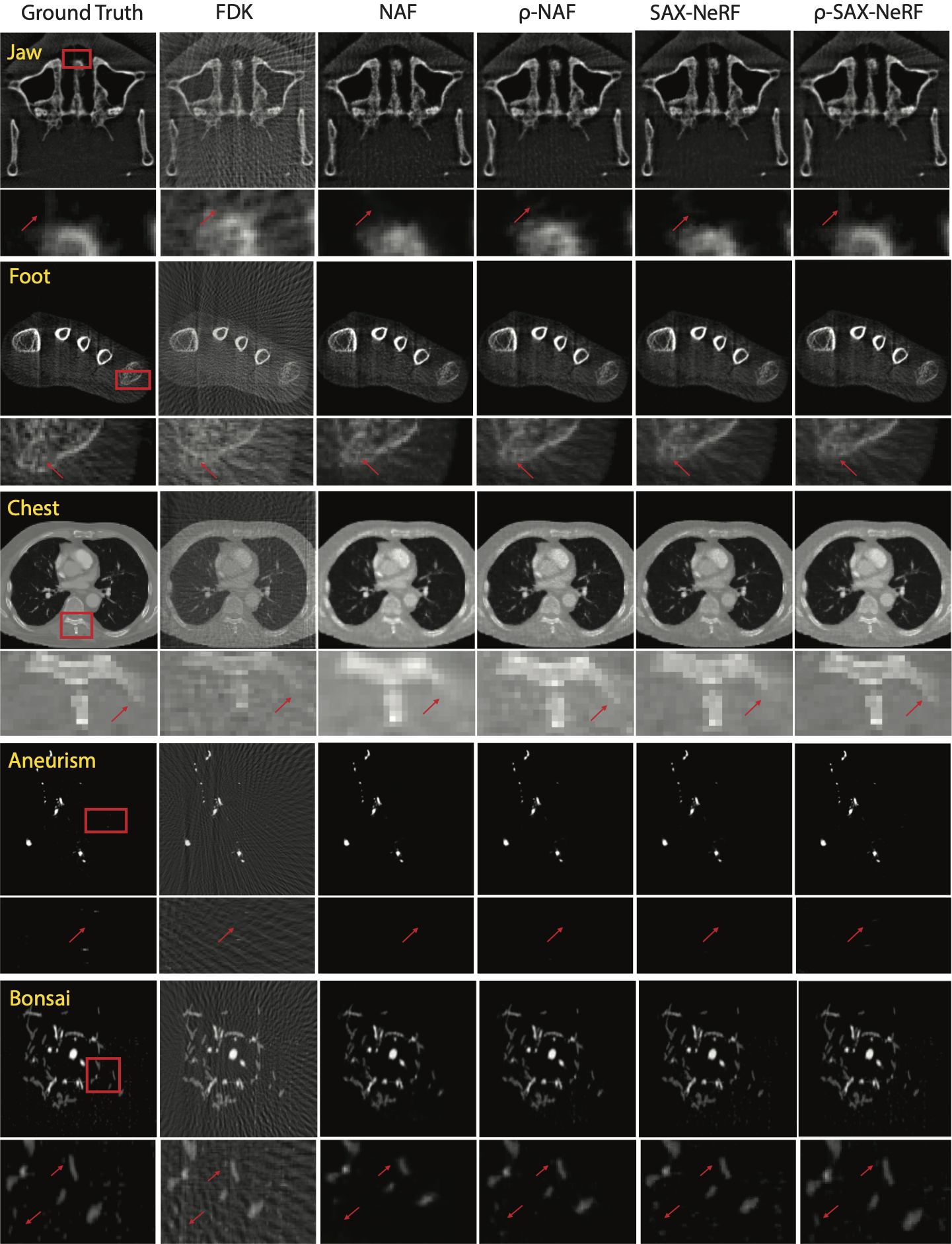}
   \caption{Visualization of some representative CT reconstruction results for Jaw, Foot, Chest, Aneurism, and Bonsai (from top to bottom). Zoomed region-of-interests are also provided.}
   \label{fig:recosntruction}
\end{figure*}

\textbf{3D CT Reconstruction:}
\cref{tab:reconstruction} and~\cref{fig:recosntruction} provide quantitative and visual comparisons of 3D CT reconstruction across baseline and $\rho$-NeRF-enhanced models. 
The $\rho$-based models consistently outperform their counterparts across all cases. Specifically, SSIM values improve from 0.93 to 0.95 in foot reconstructions when comparing NAF and $\rho$-NAF, also it surpasses SAX-NeRF-based models while maintaining a negligible computational cost, which will be discussed in~\cref{sec:ablation}. In chest scans, $\rho$-NAF achieves a PSNR of 35.07 dB, compared to 34.46 dB with SAX-NeRF, underscoring the positive impact of leveraging attenuation priors. These results validate that $\rho$-NeRF effectively reduces artifacts and yields more precise reconstructions, especially in sparse-view conditions, emphasizing the benefits of refined attenuation initialization.

\begin{figure*}
  \centering
   \includegraphics[width=0.9\linewidth]{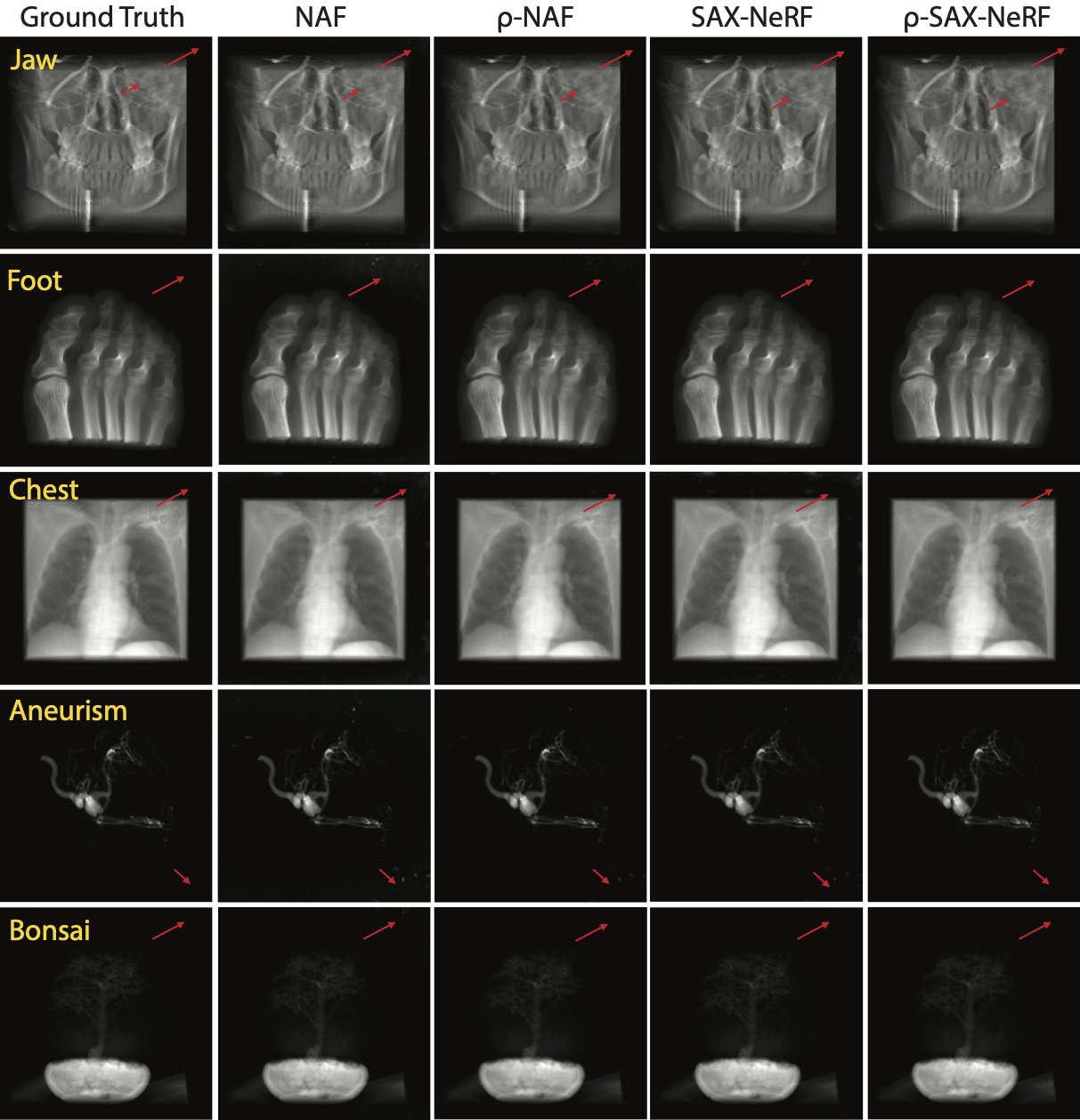}
   \caption{Visualization of some representative novel view results for Jaw. Foot, Chest, Aneurism, and Bonsai (from top to bottom).}
   \label{fig:projection}
\end{figure*}

\textbf{Novel View Synthesis:} 
In~\cref{tab:projection}, quantitative evaluation results of novel view synthesis (NVS)  show that $\rho$-NAF and $\rho$-SAX-NeRF provide substantial improvements over approaches adopting only spatial location as inputs, and promising the effectiveness of attenuation priors mechanism.
For instance, $\rho$-SAX-NeRF achieves a PSNR of 47.66 dB and an SSIM of 0.9996 on chest views, compared to SAX-NeRF’s 47.41 dB and 0.9994, respectively. 
~\cref{fig:projection} visually demonstrates the sharper anatomical details rendered by $\rho$-based models, effectively capturing fine structures. This performance, combined with supplementary results, highlights the capability of $\rho$-NeRF in delivering visually accurate and consistent NVS results across various anatomical regions.

\begin{figure}[t]
  \centering
   \includegraphics[width=0.9\linewidth]{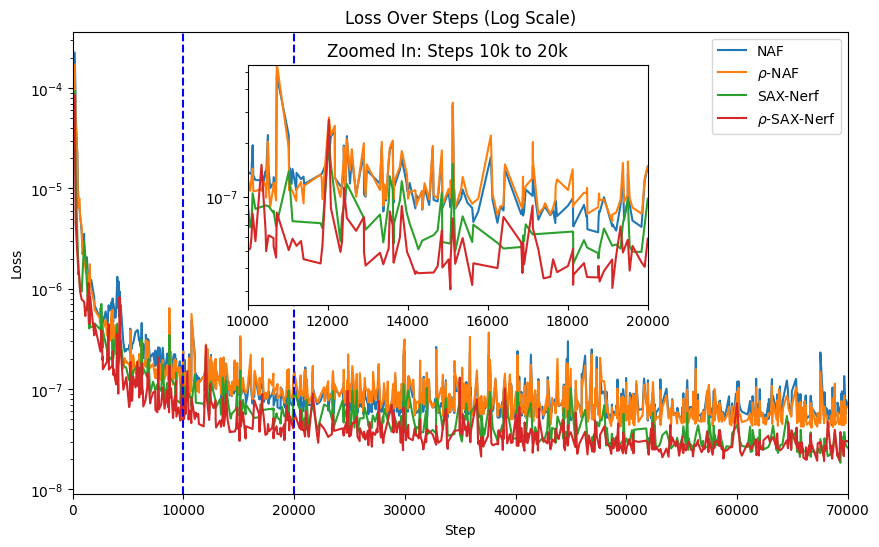}
   \caption{Model convergence with respect to training steps. }
   \label{fig:convergence}
\end{figure}

\subsection{Ablation Study}
\label{sec:ablation}

\textbf{Attenuation Initialization Analysis:}
FDK is our primary choice for initialization in the pipeline due to its ability to retain crucial attenuation information, providing a solid baseline for the $\rho$-based models. We evaluate three traditional initialization algorithms—FDK, ASD-POCS, and CGLS—to estimate initial attenuation coefficients for the~$\rho$-NeRF models. As shown in~\cref{tab:initial}, FDK consistently achieves high SSIM and PSNR scores, confirming its effectiveness for initializing attenuation values. FDK, being a physics-based algorithm, preserves more attenuation details, whereas iterative methods with TV-regularization, such as ASD-POCS and CGLS, may result in overly smooth reconstructions, potentially losing physical details. For these reasons, we primarily use FDK for initialization in our main experiments.

\begin{table}[ht]
\centering
\begin{tblr}{
  vline{2} = {-}{},
  hline{1,7} = {-}{0.08em},
  hline{2} = {-}{0.05em},
}
$\rho$-NAF & FDK          & ASD-POCS     & CGLS         \\
Jaw                     & \textbf{35.23/0.9512} & 34.48/0.9420 & 34.87/0.9466 \\
Foot                    & 32.84/0.9506 & \textbf{32.84/0.9507} & 32.20/0.9408 \\
Chest                   & \textbf{35.07/0.9733} & 33.39/0.9689 & 33.13/0.9654 \\
Aneurism                & 41.46/0.9956 & \textbf{41.52/0.9973} & 39.43/0.9928 \\
Bonsai                  & \textbf{36.11/0.9751} & 35.31/0.9656 & 34.23/0.9627 
\end{tblr}
\caption{Ablation study results of attenuation initialization algorithms in terms of PSNR/SSIM, where ~\textbf{bold} fonts indicate the best results.}
\label{tab:initial}
\end{table}

\textbf{Model Complexity Analysis:} 
~\cref{tab:complexity} compares the computational cost and parameter size for baseline models (NAF and SAX-NeRF) against their~$\rho$-based counterparts ($\rho$-NAF and $\rho$-SAX-NeRF). The $\rho$-based models introduce minimal increases in computational cost, with only 0.19 giga multiply-add operations per second (GMACs) increase. Parameter counts remain nearly identical across versions, with a negligible increase of less than 0.1\%. Combined the results in ~\cref{tab:reconstruction} and ~\cref{tab:projection}, one can see that our $\rho$-based models achieve higher performance with minimal increase in computational cost.

\begin{table}
    \centering
    \begin{tblr}{
      hline{1,4} = {-}{0.08em},
      hline{2} = {-}{0.05em},
      colsep=2pt, %
      stretch=0.9 %
    }
               & NAF       & $\rho$-NAF   & SAX-NeRF  & $\rho$-SAX-NeRF \\
    GMACs      & 0.7559    & 0.9463       & 12.6273   & 12.8177         \\
    Params (M) & 14.2667   & 14.2677      & 14.3252   & 14.3262         \\
    \end{tblr}
    \caption{Comparison analysis of GMACs and parameters for different models.}
    \label{tab:complexity}
\end{table}

\textbf{Convergence Analysis:} 
~\cref{fig:convergence} shows the loss over training steps for four models: NAF, $\rho$-NAF, SAX-NeRF, and $\rho$-SAX-NeRF. The vertical axis represents the loss in log scale, and the horizontal  axis indicates the training steps up to 70,000. The inset zooms highlight the behavior of each model. Models that incorporate attenuation priors ($\rho$-NAF and $\rho$-SAX-NeRF) exhibit faster convergence and maintain lower loss values throughout the training compared to the counterparts without these priors (NAF and SAX-NeRF). This suggests that leveraging attenuation priors accelerates the convergence process, allowing the models to reach a stable and lower loss more quickly than those without the priors.

\section{Conclusion}
\label{sec:conclusion}

This work introduces~$\rho$-NeRF, a self-supervised neural framework designed for high-fidelity 3D CT reconstruction from sparse-view X-ray data. By leveraging attenuation priors with a continuous 4D representation,~$\rho$-NeRF advances the performance of CT imaging while reducing reliance on dense data or supervised learning. Key contributions include a novel input schema that integrates initialized attenuation priors, effective feature encoding methods, and robust interpolation techniques. Evaluations on the X3D dataset validate~$\rho$-NeRF’s state-of-the-art performance in novel view synthesis and CT reconstruction. This framework not only sets a new benchmark in efficiency and accuracy for sparse-view CT applications but also offers potential for adaptation across a range of Nerf-based methods on CT reconstruction, contributing to advancements in medical and industrial CT imaging.

{
    \small
    \bibliographystyle{ieeenat_fullname}
    \bibliography{main}
}

\clearpage
\setcounter{page}{1}
\maketitlesupplementary

\section{Attenuation Initialization Analysis}
\label{sec:supp_initial}

We present additional experiments to extend the analysis of attenuation initialization strategies discussed in~\cref{sec:ablation} of the main paper. These experiments focus on two key aspects: the role of pre-initialization algorithms and the impact of different interpolation methods. By providing further insights into these components with both quantitative results and qualitative visualizations, we aim to highlight their impacts on the overall CT reconstruction quality and validate the robustness of leveraging attenuation priors.

\begin{figure*}
  \centering
   \includegraphics[width=0.9\linewidth]{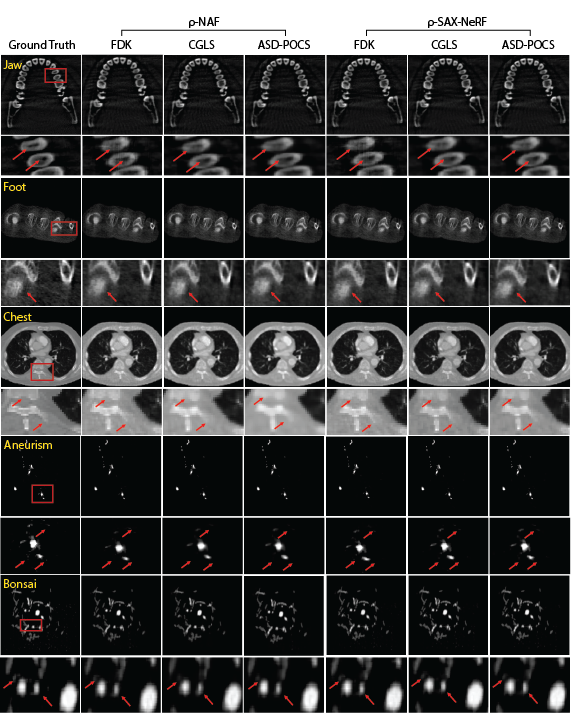}
   \caption{Visualization of representative results from different pre-initalization algorithms for Jaw, Foot, Chest, Aneurism, and Bonsai (from top to bottom). Zoomed region-of-interests are also provided.}
   \label{fig:initial1}
\end{figure*}

\textbf{Pre-initialization Algorithms Analysis:}
The first part of this analysis, as demonstrated in~\cref{tab:initial1} and~\cref{fig:initial1} , evaluates the performance of different pre-initialization algorithms—FDK, CGLS, and ASD-POCS—using the nearest interpolation method. The findings reveal that FDK consistently provides superior initialization for both $\rho$-NAF and $\rho$-SAX-NeRF models, achieving the best balance between detail preservation and reconstruction quality. While the ASD-POCS excels in some specific scenarios, its smoothing tendency often reduces fine detail accuracy, making FDK the preferred choice for reliable initialization.

\begin{table*}
\centering
\begin{tblr}{
  vline{2,5} = {-}{},
  hline{1,8} = {-}{0.08em},
  hline{2} = {2-4, 5-7}{0.05em},
}
match\_mode &          &  $\rho$-NAF     &              &              & $\rho$-SAX-NeRF         &     \\
= Nearest       & FDK    & CGL    & SD-POCS & FDK    & CGLS    & ASD-POCS \\
\hline
Jaw                  & 35.23/0.9512    & 34.87/0.9466     & 34.48/0.9420      & \textbf{35.25/0.9522}    & 35.20/0.9509     & 34.93/0.9468      \\
Foot                 & 32.84/0.9506    & 32.20/0.9408     & \textbf{32.84/0.9507}      & 32.72/0.9481    & 32.23/0.9411     & 32.29/0.9422      \\
Chest                & \textbf{35.07/0.9733}    & 33.13/0.9654     & 33.39/0.9689      & 34.81/0.9747    & 34.76/0.9746     & 34.82/0.9747      \\
Aneurism             & 39.43/0.9928    & 39.43/0.9928     & \textbf{41.52/0.9973}      & 41.41/0.9966    & 41.35/0.9960     & 41.00/0.9958      \\
Bonsai               & 34.79/0.9705    & 34.23/0.9627     & 35.31/0.9656      & \textbf{36.85/0.9798}    & 36.70/0.9780     & 36.79/0.9771      \\
\end{tblr}
\caption{Quantitative comparison of CT reconstruction results with different pre-initialization algorithms using the nearest interpolation method in terms of PSNR/SSIM, where~\textbf{bold} fonts indicate the best results.}
\label{tab:initial1}
\end{table*}

\begin{figure*}
  \centering
   \includegraphics[width=0.9\linewidth]{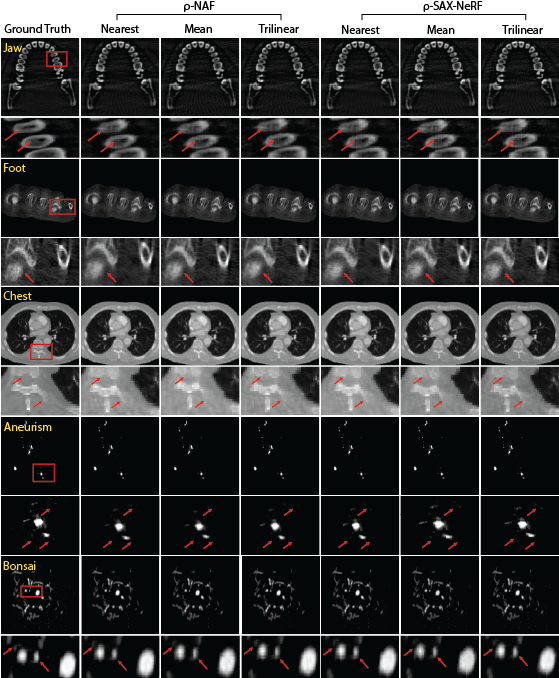}
   \caption{Visualization of representative resutls from different interpretation methods for Jaw, Foot, Chest, Aneurism, and Bonsai (from top to bottom). Zoomed region-of-interests are also provided.}
   \label{fig:initial2}
\end{figure*}

\textbf{Interpolation Methods Analysis:}
The second part of this analysis, shown in~\cref{tab:initial2} and~\cref{fig:initial2}, investigates the impact of interpolation methods—Nearest, Mean, and Trilinear—on reconstruction performance when using FDK as the pre-initialization algorithm. These experiments evaluate $\rho$-NAF and $\rho$-SAX-NeRF models across the same five cases. Nearest interpolation generally demonstrates better performance due to its ability to preserve sharp boundaries, particularly in simpler cases like Foot and Chest. Trilinear interpolation, however, provides competitive or superior results in more complex scenarios such as Aneurism and Bonsai, benefiting from smoother transitions across voxel grids. This analysis underscores the adaptability of interpolation techniques and their critical role in optimizing reconstruction outcomes.

\begin{table*}
\centering
\begin{tblr}{
  vline{2,5} = {-}{},
  hline{1,8} = {-}{0.08em},
  hline{2} = {2-4, 5-7}{0.05em},
}
image\_init &         &   $\rho$-NAF &         &           &   $\rho$-SAX-NeRF         &   \\
= FDK            & Nearest & Mean & Trilinear & Nearest & Mean & Trilinear \\
\hline
Jaw                     & 35.23/0.9512     & 34.53/0.9421  & 34.85/0.9460       & \textbf{35.25/0.9522}     & 34.95/0.9477  & 34.89/0.9468       \\
Foot                    & \textbf{32.84/0.9506}     & 32.40/0.9488  & 32.56/0.9498       & 32.72/0.9481     & 32.53/0.9480  & 32.73/0.9507       \\
Chest                   & \textbf{35.07/0.9733}     & 32.59/0.9605  & 33.49/0.9644       & 34.81/0.9747     & 34.71/0.9741  & 34.80/0.9749       \\
Aneurism                & 39.43/0.9928     & 36.98/0.9899  & 38.59/0.9909       & 41.41/0.9966     & 40.67/0.9956  & \textbf{42.05/0.9962}       \\
Bonsai                  & 34.79/0.9705     & 35.06/0.9710  & 35.05/0.9720       & 36.85/0.9798     & 36.55/0.9782  & \textbf{36.95/0.9805}       \\
\end{tblr}
\caption{Quantitative comparison of representative results with different interpolation methods using the FDK pre-initialization algorithm in terms of PSNR/SSIM, where ~\textbf{bold} fonts indicate the best results.}
\label{tab:initial2}
\end{table*}

\end{document}